\documentclass[aps,
prb,
twocolumn,
amsmath,
amssymb,
superscriptaddress,
reprint,
10pt,
floatfix
]{revtex4-2}

\usepackage{url}
\usepackage{txfonts}
\usepackage{float}
\usepackage{amsmath, amssymb}
\usepackage{pstricks}
\usepackage{graphicx}
\usepackage{tikz}
\usepackage[T1]{fontenc}
\usepackage{color}
\usepackage{chemformula}
\usepackage[colorlinks=true,breaklinks=true,allcolors=teal]{hyperref}

\usepackage{tabularx}
\usepackage{multirow}
\usepackage{array}

\usepackage{physics}
\usepackage{xparse}
\usepackage[version=4]{mhchem}
\usepackage[normalem]{ulem}

\makeatletter
\let\cat@comma@active\@empty
\makeatother

\begin{document}

\title{Excitonic quasi-condensation in a mass-asymmetric electron-hole biwire}

\author{Gautam Shah}

\author{Vinod Ashokan}
\email{ashokanv@nitj.ac.in}
\affiliation{%
Computational Quantum Many-Body Physics Lab,
Department of Physics, Dr.\ B.\ R.\ Ambedkar National Institute of Technology, Jalandhar, Punjab - 144008, India
}
\author{N.\ D.\ Drummond}
\affiliation{Department of Physics, Lancaster University, Lancaster LA1 4YB, United Kingdom}%

\author{K.\ N.\ Pathak}
\affiliation{%
Centre for Advanced Study in Physics, Panjab University, Chandigarh - 160014, India
}%

\begin{abstract}
We study a mass-asymmetric one-dimensional electron-hole biwire system at zero temperature using the diffusion quantum Monte Carlo method. We characterize electron-hole pairing and long-distance pair coherence through the pair correlation function and the projected two-body density matrix over a broad range of carrier densities $r_{\rm s}$ and interwire separations $d$. We identify distinct correlation regimes ranging from a weakly correlated two-component electron-hole plasma to a quasi-condensate excitonic regime and a strongly correlated Wigner-like regime. To understand long-distance pair coherence, we analyze the algebraic decay of the projected two-body density matrix and extract the corresponding power-law exponent $\alpha$ which is less than 2. Strong electron-hole Coulomb correlations allow quasi-excitonic coherence to persist to high densities, down to effective $r_{\rm s} = 0.3$ a.u.\ in addition to low densities. For a representative parameter set, we further find that mass asymmetry suppresses long-distance pair coherence while enhancing short-range electron-hole pairing. Our results provide a microscopic characterization of competing pairing and spatial ordering correlations in mass-asymmetric one-dimensional electron-hole systems.
\end{abstract}

\maketitle

\section{Introduction}

The formation of bound electron-hole pairs and their collective behavior lie at the heart of excitonic physics~\cite{Nandi2012}. When electrons and holes bind into charge-neutral composite bosons, they may develop macroscopic quantum coherence, analogous to Bose-Einstein condensation~\cite{Eisenstein2004} or Cooper pairing in superconductors~\cite{Anirban2025,Kavokin2016}. Beyond its fundamental significance, excitonic ordering has attracted considerable interest for its potential applications in dissipationless transport~\cite{Zhang2022} and quantum-coherent devices~\cite{Peotta2011,Liu2017}. However, identifying and characterizing excitonic coherence in equilibrium many-body systems remains challenging, particularly in regimes where Coulomb correlations are strong. Low-dimensional electron-hole systems provide a promising route towards enhancing interaction effects~\cite{Cazalila,Guan}. Reduced dimensionality suppresses screening and strengthens Coulomb correlations, favoring exciton formation even at higher carrier densities~\cite{Nagaosa1993}. In one dimension (1D), true long-range order is absent in the thermodynamic limit because of strong quantum fluctuations. Nevertheless, strong electron-hole pairing can produce algebraically decaying pair correlations, even in the absence of true long-range order~\cite{abergel2015excitonic}. In particular, quasi-one-dimensional superconductors have been shown to exhibit critical temperatures exceeding their bulk counterparts due to quantum confinement and subband effects~\cite{Bianconi1998,Shanenko2006}. The underlying physics can be realized in a variety of semiconductor and nanoscale platforms, including GaAs-based heterostructures~\cite{Steinberg2006,DasGupta2011}, graphene nanoribbons~\cite{ribbons}, and core-shell nanowires~\cite{Ganjipour2012}. Despite substantial theoretical interest, a quantitatively reliable description of excitonic coherence in 1D  systems remains incomplete. Previous theoretical studies have largely relied on mean-field, bosonization~\cite{wermen}, or self-consistent perturbative approaches~\cite{Saini2004,Moudgil2010}, which can become insufficient when strong coupling and bound-state formation are important. Quantum Monte Carlo studies have also investigated electron-hole (e-h) bilayers, including their excitonic correlations and the effects of mass asymmetry~\cite{Palo2002, maezono2013, RO2016}. However, the corresponding correlations in a 1D e-h biwire remain to be characterized.

In this paper, we address these challenges using diffusion Monte Carlo (DMC) simulations of an electron-hole biwire (EHBW) system with realistic mass-asymmetry over a broad range of carrier densities and interwire separations. We compute spin-resolved pair correlation functions and the projected two-body density matrix (TBDM) to characterize both local pairing and long-distance pair coherence. On this basis, we identify distinct regimes of the EHBW system dominated by weak electron-hole correlations, excitonic pairing and Wigner-like spatial ordering. We further examine the effects of mass-asymmetry and find that increasing carrier-mass imbalance enhances short-range electron-hole pairing while reducing long-distance pair coherence.

DMC is a stable method that is known to yield extremely accurate results for 1D homogeneous electron gases~\cite{lee2011,Ashokan2018}. This high accuracy arises because, in 1D, the fermionic nodal structure is exactly determined by particle ordering, rendering fixed-node DMC exact. However, for realistic finite-width interactions ergodicity issues may arise, as opposite-spin particles can only rarely exchange positions; these effects are avoided in the strictly 1D model considered in the present work.

\section{Model}

We consider an EHBW system consisting of two parallel, mass-asymmetric, infinitely thin quantum wires separated by a distance \(d\), with $N_{\rm e} \equiv N$ spin-up electrons confined to one wire and $N_{\rm h} \equiv N$ spin-down holes to the other in a cell of length $L$ subject to 1D-periodic boundary conditions. The effective masses of electrons \(m_{\rm e}^\ast\) and holes \(m_{\rm h}^\ast\) are unequal (\(m_{\rm e}^\ast \ne m_{\rm h}^\ast\)). The dimensionless density parameter is defined as \( r_{\rm s} = {1}/{(2 n a_{\rm B}^\ast)} \), where $n=N/L$ is the linear carrier density in each wire. In SI units, the effective Bohr radius and Hartree energy are given by \(a_{\rm B}^\ast = {4\pi\epsilon \, \hbar^2}/{(m_{\rm e}^\ast e^2)} \), \(1\,\mathrm{Ha}^\ast = {e^2}/{(4\pi\epsilon \, a_{\rm B}^\ast)} \). Henceforth, we use effective Hartree atomic units, where \(\hbar = |e| = m_{\rm e}^\ast = 4\pi\epsilon = 1\) a.u., where $\epsilon$ is the static permittivity of the surrounding medium.

The Hamiltonian of the EHBW system consisting of  \(N_{\rm e}\)= \(N_{\rm h} = N\) particles per wire is expressed as
\begin{equation}
\begin{aligned}
	\hat{H} = &-
	\frac{1}{2} \sum_{i=1}^{N_{\rm e}} \frac{\partial^2}{\partial x_{i,{\rm e}}^2}
	- \frac{1}{2\sigma} \sum_{i=1}^{N_{\rm h}} \frac{\partial^2}{\partial x_{i,{\rm h}}^2} \nonumber \\
	&+ \sum_{i<j}^{N_{\rm e}} V(|x_{i,{\rm e}}-x_{j,{\rm e}}|,0)
	+ \sum_{i<j}^{N_{\rm h}} V(|x_{i,{\rm h}}-x_{j,{\rm h}}|,0) \nonumber \\
	&- \sum_{i=1}^{N_{\rm e}} \sum_{j=1}^{N_{\rm h}}
	V(|x_{i,{\rm e}}-x_{j,{\rm h}}|,d)
	 +\frac{N_{\rm e}+N_{\rm h}}{2} v_{\rm M} + V_{\mathrm{bb}},
\end{aligned}
\end{equation}
where \(x_{i,\alpha}\) denotes the position of particle \(i\) in wire \(\alpha\), where \(\alpha = {\rm e}\) for electrons and \(\alpha = {\rm h}\) for holes. $V(x,z)$ is the 1D Ewald interaction between particles with in-wire separation $x$ and interwire separation $z$, describing the Coulomb interaction between electrons in one wire and holes in the other and $v_{\rm M}$ is the Madelung constant~\cite{Saunders1994}, which is the sum of pairwise interactions of a particular particle with all its periodic images. $V_{\mathrm{bb}}=-[N_{\rm e}N_{\rm h}/(2L)] \int_{-L/2}^{L/2} V(x,z) \, dx$ is a constant energy contribution arising from the interaction between the neutralizing backgrounds of the two wires.
The electron mass is $m_{\rm e}^\ast = 1$ a.u.\ and the hole mass is $m_{\rm h}^\ast = \sigma m_{\rm e}^\ast$, where $\sigma$ is the hole-to-electron mass ratio, which is set to $7$ in this study. This value is representative of the mass asymmetry encountered in GaAs-based electron-hole systems~\cite{Steinberg2006}. It is well established that in an infinitely thin 1D wire, the ground-state many-body wave function of fermions with Coulomb interactions exhibits nodes at every particle coalescence point, regardless of the spin configuration~\cite{lee2011}. Consequently, the paramagnetic and ferromagnetic states are exactly degenerate, and the Lieb-Mattis theorem~\cite{liebPhysRev.125.164} does not apply. The ground-state energy therefore depends solely on the particle density and not on the degree of spin polarization. 

\section{Computational method}
The DMC method as implemented in \textsc{casino} code~\cite{Needs2020} was used to study the EHBW system. DMC is a projector technique in which the ground state properties are obtained by evolving the many-body wave function in imaginary time towards the lowest energy solution~\cite{ceperley1980,foulkes2001}. The fermionic antisymmetry is enforced by fixing the wave-function nodes at the coalescence points~\cite{Anderson,Anderson2}, which are also the nodes of the trial wave function $\Psi_{\rm T}$. This constraint eliminates the fermion sign problem by restricting the random walks to regions of fixed sign, ensuring the stable projection onto the lowest energy state. The trial wave function $\Psi_{\rm T}$ contains a set of variational parameters that are optimized independently for each coupling parameter $r_{\rm s}$ and interwire spacing $d$ by first minimizing the mean absolute deviation of the local energies from their median~\cite{Needs2020}, and then refining them using linear least-squares energy minimization~\cite{Umrigar}.

Following the established variational Monte Carlo (VMC) study of an electron-electron (e-e) biwire system by Sharma \textit{et al.}~\cite{rajesh2021}, we employ a Slater-Jastrow-backflow trial wave function $\Psi_{\rm T}$ for our EHBW system. The Slater determinants are constructed separately for electrons and holes confined to different wires.

The Slater-Jastrow-backflow trial wave function we used is of the form
\begin{align}
\Psi_{\rm T} = D\!\left(\phi^{\uparrow}_{\rm e}(x({\bf R}))\vphantom{\phi^{\downarrow}_{\rm h}(x)}\right)\,
         D\!\left(\phi^{\downarrow}_{\rm h}(x({\bf R}))\right)\,
         e^{J({\bf R})},
\end{align}
where \(\phi^{\uparrow}_{\rm e}\) and \(\phi^{\downarrow}_{\rm h}\) denote the spin-up and spin-down orbitals of electrons and holes, respectively. \(D\) is a Slater determinant and \(e^{J({\bf R})}\) is a Jastrow factor, which accounts for the correlations between the charge carriers within as well as between the wires. The Slater determinants for both species are constructed from plane-wave orbitals \(\phi(x) = e^{ikx}\). The orbitals are evaluated at quasiparticle coordinates $x$, which are functions of all the electron and hole coordinates ${\bf R}$ via a backflow transformation. In a finite system, the first $N_{\rm e}$ and $N_{\rm h}$ plane-wave orbitals are occupied, irrespective of their relation to the infinite-system Fermi wave vector \(k_{\mathrm{F}} = \pi/2r_{s}\). 

In our calculations, we employ a Drummond-Towler-Needs Jastrow factor~\cite{drummond2004} specifically tailored for the EHBW system. It consists of three components. First a two-body polynomial ($u$) term, expanded up to eighth order, which accounts for e-e, hole-hole (h-h), and e-h correlations separately. The 1D analogs of the Kato cusp conditions~\cite{Kato1957} are enforced for e-e and h-h pairs within the Jastrow factor. For e-h pairs, the cusp constraint is implemented as a short-range condition ensuring the correct local behavior of the wave function at finite separations. To provide an additional smooth and flexible description of short-range correlations across various interwire separation $d$, we introduce a `quasicusp' Jastrow term $Q$, characterized by a single optimizable cutoff length. While the calculations are performed at finite $d$, the strictly 1D model develops a divergent attractive interaction as $d \to 0$, and thus becomes unphysical in this limit. Finally, we add a plane-wave ($p$) term, represented by 20 independent reciprocal-lattice vectors along the wire axis, with all Fourier coefficients optimized to capture long-range correlations.

To go beyond two-body correlations, we incorporate a backflow transformation~\cite{backflow2006}, in which the particle coordinates in the Slater determinants are replaced by quasiparticle coordinates defined through polynomial expansions in the particle separation up to eighth order.

The primary role of backflow is to provide a compact description of many-body correlation effects. In the present work, we include a single two-body backflow ($\eta$) channel with spin dependence, which provides a compact and efficient description of higher-order (effective three-body) correlations. This backflow parameterization does not alter the nodal structure which is already exact for infinitely thin 1D wires~\cite{backflow2006} but improves the variational flexibility of the trial wave function.

The translational average of the TBDM for e-h pairs is
\begin{equation}
\rho^{(2)}_{\rm eh}({x}) = \frac{N_{\rm e} N_{\rm h} }{L^2} 
\frac{ \int |\Psi(\mathbf{R})|^2 \, 
\frac{ \Psi({x_{\rm e}} + {x}, {x_{\rm h}} + {x}) }{ \Psi(x_{\rm e}, x_{\rm h}) } \, d\mathbf{R} }
{ \int |\Psi(\mathbf{R})|^2 \, d\mathbf{R} }.
\end{equation}
In a translationally invariant system, off-diagonal long-range order is associated with a nonzero asymptotic value of the TBDM~\cite{Yang1962}. In the thermodynamic limit of a 1D system at zero temperature, however, true condensate order is absent. We therefore characterize excitonic pair coherence through the long-distance decay of the projected e-h TBDM $c(x)$. 

\begin{equation} c(x)= N \frac{ \int |\Psi(\mathbf{R})|^2 \left[ \begin{array}{l} \frac{ \Psi({x_{\rm e}} + {x}, {x_{\rm h}} + {x}) }{ \Psi(x_{\rm e}, x_{\rm h}) } \\ \qquad {} - \frac{ \Psi({x_{\rm e}} + {x}, {x_{\rm h}}) }{ \Psi(x_{\rm e}, x_{\rm h}) } \frac{ \Psi({x_{\rm e}}, {x_{\rm h}} + {x}) }{ \Psi(x_{\rm e}, x_{\rm h}) } \end{array} \right] \, d\mathbf{R} }
{ \int |\Psi(\mathbf{R})|^2 \, d\mathbf{R} },
\end{equation}
introduced in Ref.~\onlinecite{astra2005}. To quantify this behavior, we fit the long-distance data to an algebraic form
\begin{equation}
 c(x)=\sum_{j}\frac{A}{|x-jL|^\alpha}, 
\end{equation}
where $\alpha$ is the decay exponent characterizing the  algebraic power-law behavior indicative of finite size quasi-condensate and $L$ is the simulation cell length. The fits are performed over $L/4 \lesssim x \le L/2$. If $\alpha$ turns out to be nearly zero, one will have full condensate.

The translational average of the pair correlation function (PCF) has also been computed:
\begin{equation}
g_{\alpha\alpha'}(x)=\frac{L
\int |\Psi(\mathbf{R})|^2
\sum_{\substack{i,j}}
\delta\,\big(x_{i,\alpha}-x_{j,\alpha'}-x\big)\,
d\mathbf{R}
}
{N_{\alpha}N_{\alpha'}\int |\Psi(\mathbf{R})|^2 \, d\mathbf{R}},
\end{equation}
where $x_{i,\alpha}$ denotes the position of the $i$-th particle of species $\alpha$.

The PCF measures the probability of finding a particle at a distance $x$ from a reference particle at the origin.

Our results were obtained with \(N_{\rm e} = 21\) electrons and $N_{\rm h} = 21$ holes (42 particles in total). To assess finite-size effects; see Appendix~\ref{Appendix A}, we simulated \(N=61\) particles of each species (122 particles in total). The PCFs remain unchanged within statistical uncertainty, while the finite-size quasi-excitonic behavior of the projected TBDM is analyzed using power-law fits; see Appendix~\ref{Appendix B}, Table~\ref{powerlaw}. A target population of 510 walkers was employed and we found that the DMC energy remains unchanged on increasing the population further, so population-control bias can be neglected. In the DMC calculations we employed a density-dependent time step $\tau=0.008\,r_{\rm s}^2$~\cite{lee2011} chosen such that the root-mean-square diffusion distance of an electron in a single step remains much smaller than the mean interparticle spacing. This criterion ensures stable and accurate imaginary-time propagation across the density range considered. Convergence with respect to the time step was explicitly tested at ($r_{\rm s} = 1$ a.u.\ and $d = 0.05$ a.u.), where time-step errors are expected to be most pronounced, by repeating calculations with a tenfold smaller time step. No statistically significant differences were observed in the resulting expectation values (see Appendix~\ref{Appendix C}), and all reported results are therefore considered to be converged with respect to~$\tau$.

\section{Results and discussion}

\begin{figure}[tbp!]
	\centering
	\includegraphics[width=0.95\columnwidth]{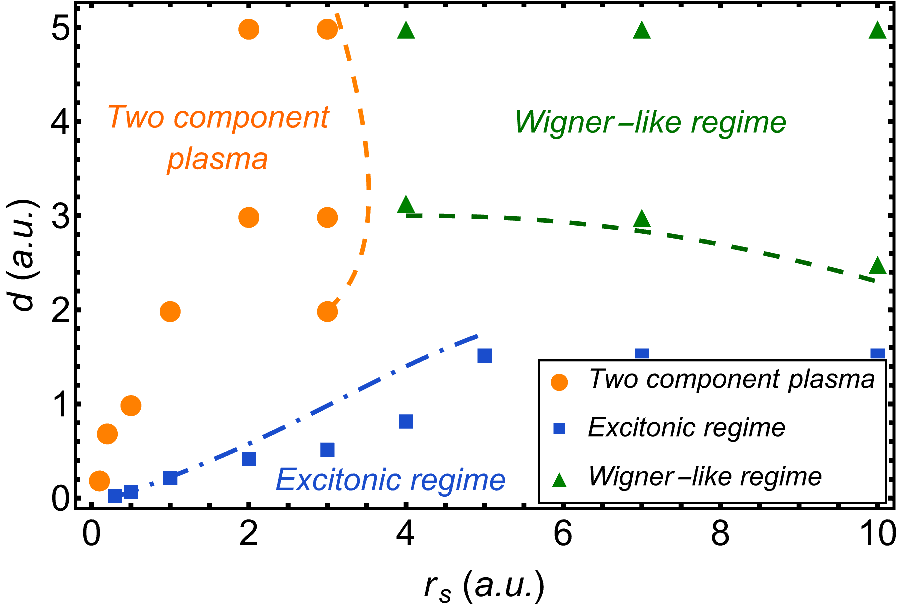}
	\caption{Ground-state correlation-regime diagram of the mass-asymmetric EHBW system in the $(r_{\rm s},d)$ plane. Orange circle, blue square, and green triangles denote the two-component (2C) plasma, excitonic, and Wigner-like regimes, respectively, identified from the PCFs, SSFs, and long-distance behavior of the e-h two-body density matrix. The dashed curves are guides to the eye indicating the approximate crossover trends and are not intended as sharply defined phase boundaries.}
	\label{phasediagram}
\end{figure}

Figure~\ref{phasediagram} summarizes the ground-state correlation regimes of the mass-asymmetric EHBW system in the $(r_{\rm s},d)$ plane. We distinguish three characteristic correlation regimes based on the combined behavior of the real-space PCFs, static structure factors (SSFs), and the long-distance behavior of the projected e-h TBDM\@. In the 2C plasma regime, the interwire e-h correlations remain relatively weak and the long-distance pair coherence is strongly suppressed. In the excitonic regime, the decreasing interwire separation enhances short-range e-h correlations and gives rise to a finite-size long-distance pair-coherence signal in the TBDM\@. In the Wigner-like regime, strong intrawire Coulomb correlations produce pronounced spatial ordering, reflected in the PCFs and SSFs; see Appendix~\ref{Appendix D}. At small $r_{\rm s}$ (high density), the system is in the 2C plasma regime, except at sufficiently small $d$, where the interwire attraction becomes strong enough to produce pronounced e-h correlations. With increasing $r_{\rm s}$, the region exhibiting excitonic correlations extends to larger interwire separations, reflecting the increasing importance of Coulomb correlations. At larger $r_{\rm s}$ and sufficiently large $d$, intrawire Coulomb correlations become dominant and the system exhibits Wigner-like spatial ordering. The individual points in Fig.~\ref{phasediagram} are assigned according to the calculated correlation signatures rather than solely by their positions relative to the dashed curves. The dashed curves therefore serve only as guides to the eye summarizing the crossover trends. Since the different correlation signatures evolve continuously and need not change simultaneously, the corresponding crossover guides are not expected to meet at a unique point. Within the parameter range investigated, strong excitonic correlations and finite-size long-distance pair coherence persist to the high-density limit, reaching $r_{\rm s} = 0.3$ a.u.\ at $d = 0.01$ a.u. Thus, Fig.~\ref{phasediagram} is intended as a map of distinct correlation regimes and their crossover trends, rather than as a thermodynamic phase diagram with sharply defined phase boundaries.                

\begin{figure}[tbp!]
	\centering
	\includegraphics[width=\columnwidth]{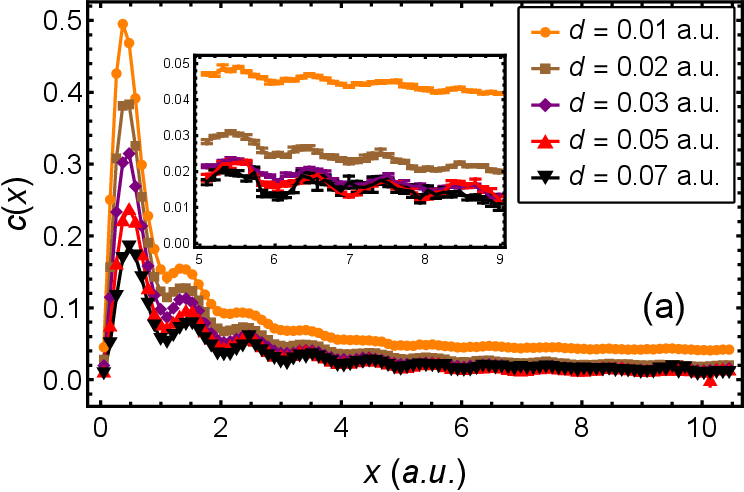}
	\par\smallskip
	\includegraphics[width=\columnwidth]{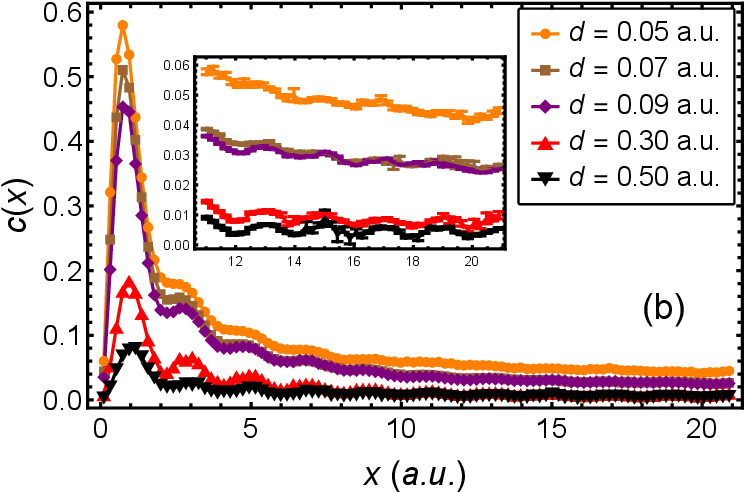}
	\par\smallskip
	\includegraphics[width=\columnwidth]{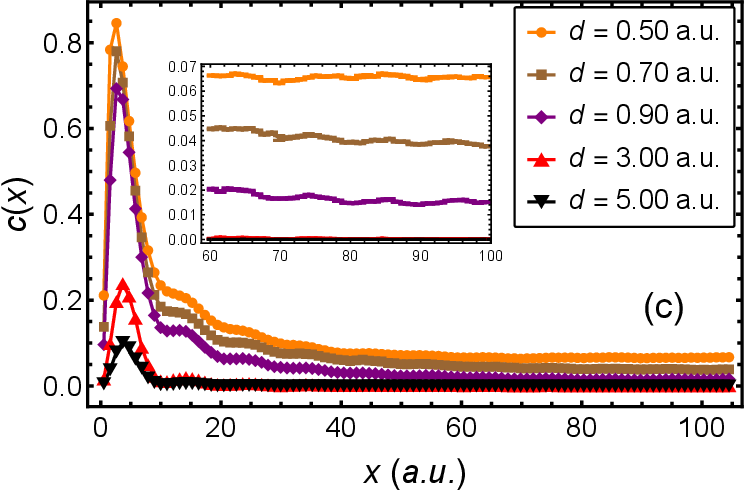}
	\caption{DMC projected TBDM $c(x)$ for the mass-asymmetric EHBW system at densities $r_{\rm s} = 0.5$, $1.0$, and $5.0$ a.u., shown in panels (a), (b), and (c), respectively, over a wide range of interwire separation $d$. The inset displays a zoomed-in view of the largest-distance data points.}\label{tdm}
\end{figure}

We examine the long-distance behavior of the projected TBDM $c(x)$ using power-law fits, as summarized in Appendix \ref{Appendix B}, Table~\ref{powerlaw} for three representative densities, $r_{\rm s} = 0.5,1.0$, and $5.0$ a.u.\ over a wide range of interwire separation $d$. The corresponding DMC results for $c(x)$ are shown in Fig.~\ref{tdm}.

At high density $r_{\rm s} = 0.5$ a.u., shown in Fig.~\ref{tdm}(a), the fitted exponent $\alpha$ varies from $1.20(1)$ to $1.72(10)$ over the interwire separations considered. The smallest exponent occurs at $d = 0.01$ a.u., where the e-h attraction is strongest among the separations studied. The plateau feature is observed up to $d = 0.05$ a.u., beyond which it becomes less apparent. At intermediate density, $r_{\rm s} = 1.0$ a.u.\ [Fig.~\ref{tdm}(b)], the fitted exponent $\alpha$ ranges from $1.47(1)$ to $1.80(18)$, and the plateau persists up to $d = 0.1$ a.u. At low density $r_{\rm s} = 5.0$ a.u.\ [Fig.~\ref{tdm}(c)], the dependence on interwire separation becomes more pronounced. The exponent increases from $1.01(3)$ at $d = 0.5$ a.u.\ to $1.41(1)$ at $d = 0.7$ a.u.\ and $1.76(3)$ at $d = 0.9$ a.u. The plateau feature remains visible up to $d = 1.0$ a.u. At larger separations, the exponent rises to $2.82(16)$ and $3.53(37)$ for $d = 3.0$ and $5.0$ a.u., respectively. The larger fitted power-law exponent indicates faster decay over the fitting interval, consistent with the weakening of the plateau behavior at large $d$. Taken together, these results show that the long-distance behavior of the projected TBDM depends on both density $r_{\rm s}$ and interwire separation $d$. Notably, the plateau feature persists over an increasingly broad range of $d$ as the density decreases (high $r_{\rm s}$). The corresponding range at $r_{\rm s} = 5.0$ a.u.\ is approximately ten times that at $r_{\rm s} = 1.0$ a.u.\ and twenty times that at $r_{\rm s} = 0.5$ a.u. The plateau accompanied by small fitted exponents at smaller $d$, is indicative of enhanced e-h pair correlations in the finite system studied.

\begin{figure*}[tbp!]
	\centering
	\begin{minipage}[t]{0.32\textwidth}
		\centering
		\includegraphics[width=\linewidth]{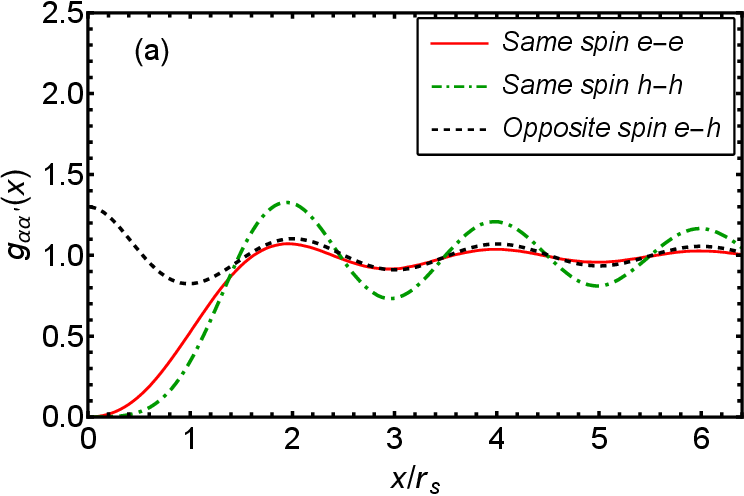}
		\par\smallskip
	\end{minipage}\hfill
	\begin{minipage}[t]{0.32\textwidth}
		\centering
		\includegraphics[width=\linewidth]{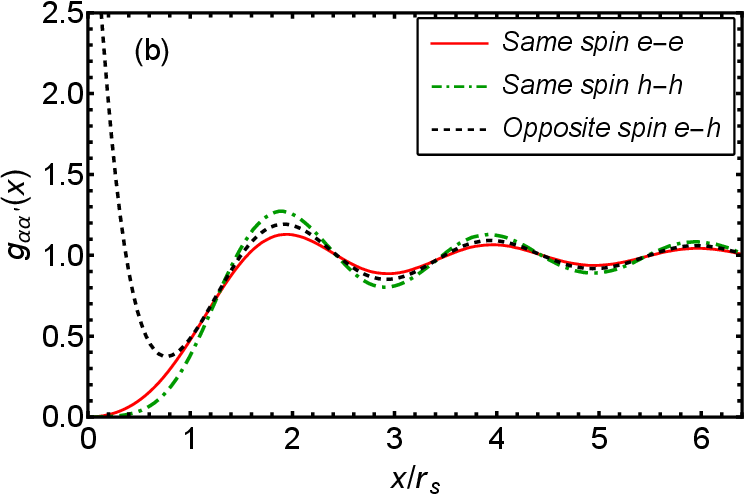}
		\par\smallskip
		
	\end{minipage}\hfill
	\begin{minipage}[t]{0.32\textwidth}
		\centering
		\includegraphics[width=\linewidth]{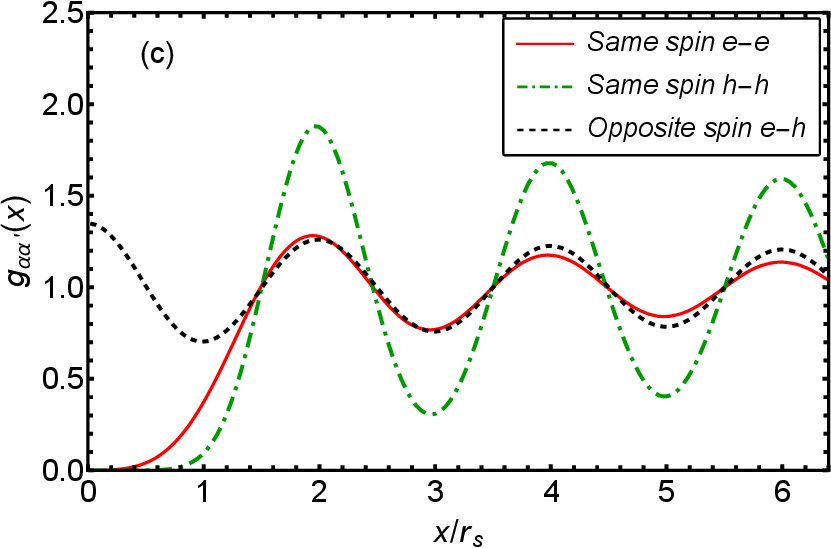}
		\par\smallskip
		
	\end{minipage}
	\caption{DMC PCF $g_{\alpha\alpha'}(x)$ as a function of
		interparticle distance $x$ for a mass asymmetric EHBW system. Panels (a)--(c) correspond to $r_{\rm s}=1$ a.u., $d=0.5$ a.u.\ (2C fluid),
		$r_{\rm s}=0.5$ a.u., $d=0.01$ a.u.\ (excitonic regime), and $r_{\rm s}=5$ a.u., $d=6$ a.u.\ (Wigner-like regime), respectively.}
	\label{pcf}
\end{figure*}

The spin-resolved PCFs $g_{\alpha\alpha'}(x)$ shown in Fig.~\ref{pcf} provide further insights into the nature of the ground state correlations obtained from DMC for an EHBW system in the three representative regimes (2C fluid regime, excitonic regime, and Wigner-like regime). The 2C fluid regime is shown in Fig.~\ref{pcf}(a) ($r_{\rm s} = 1$ a.u., $d = 0.5$ a.u.). While the same spin e-e and h-h PCFs go to zero at $x=0$, the opposite spin e-h PCF is gently peaked at $x=0$, indicating weak e-h binding. There is a difference in the peak amplitudes of the e-e and h-h PCFs, which is an artifact of the mass asymmetry between them ($m_{\rm h}^*$ = $7 m_{\rm e}^*$). The peak height in the e-h PCF is not too significant in this regime, which indicates that there is not enough bound-state formation, indicating that strong excitonic pairing has not yet developed. Figure~\ref{pcf}(b) shows the exciton regime where the peak height in e-h PCF is strongly enhanced ($r_{\rm s} = 0.5$ a.u., $d = 0.01$ a.u.), signaling strong interwire interaction. In contrast, the same spin correlations (e-e/h-h) remain suppressed at short-range while developing oscillations, with their peak amplitudes nearly superimposing with each other. The strong e-h pairing is accompanied by enhanced finite-size long-distance pair coherence and is also reflected in the  momentum density and two particle density matrix in k-space; see Appendix~\ref{Appendix E} and~\ref{Appendix F} as both species nearly approach each other. At low-density [Fig.\ref{pcf}(c)] and large separation ($r_{\rm s} = 5.0$ a.u., $d = 6$ a.u.)\ all PCFs display long-range oscillations with enhanced amplitudes, indicating an increasing degree of spatial ordering driven by intrawire repulsion. In contrast to the excitonic regime, where the dominant feature is the enhancement of the opposite e-h correlation at short-range, the low-density behavior is characterized primarily by enhanced same species correlations and long-range oscillatory structure.
The distinction between the fluid and Wigner-like regimes is subtle in this 1D system, as illustrated by Figs.~\ref{pcf}(a) and \ref{pcf}(c). In both regimes, the h-h PCF is strongly suppressed near $x = 0$, reflecting the pronounced spatial correlations of the heavier holes. The more noticeable difference is in the e-e PCF, for which the region of strong suppression around $x = 0$ is broader in the Wigner-like regime, indicating enhanced spatial correlations of the lighter electrons, which is also reflected in the finite size analysis of the PCFs presented in Appendix~\ref{Appendix A}.  This behavior is further supported by the pronounced peak at $4k_{\rm F}$ in the hole-hole SSF, which reflects the stronger spatial ordering of the heavier holes and their tendency towards localization at lower density; see Appendix~\ref{Appendix D}.

\begin{figure}[tbp!]
	\centering
	\includegraphics[width=0.91\columnwidth]{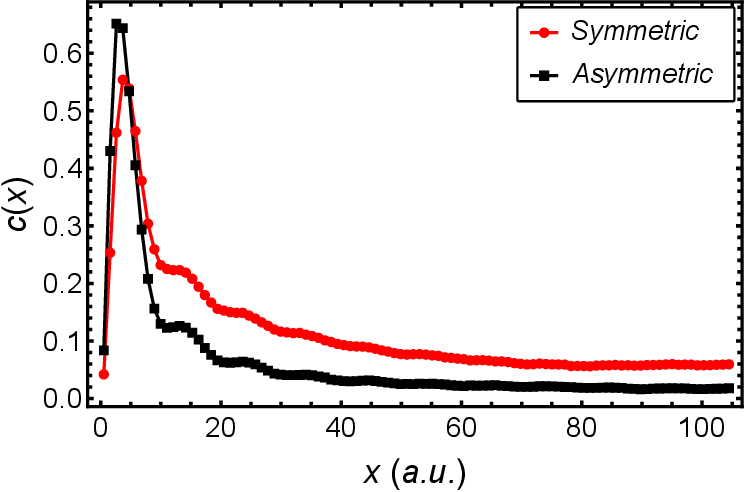}
	\caption{DMC projected TBDM $c(x)$ for the EHBW system at $r_{\rm s} = 5$ a.u.\ and interwire separation $d = 1$ a.u. Results are shown for symmetric and asymmetric mass configurations. Mass asymmetry suppresses the long-distance e-h pairing amplitude and accelerates its spatial decay relative to the symmetric case.}\label{con}
\end{figure}

To elucidate the role of mass asymmetry, we compare the projected TBDM $c(x)$ of the mass asymmetric ($m_{\rm h}^\ast = 7 m_{\rm e}^\ast$) system with the corresponding symmetric case ($m_{\rm h}^\ast = 1 m_{\rm e}^\ast$) at a fixed $r_{\rm s} = 5.0$ a.u.\ and $d = 1.0$ a.u.\ as shown in Fig.~\ref{con}. The mass asymmetric case shows a slightly higher and sharper maximum at small $x$, indicating a stronger e-h pairing amplitude. However, this enhancement is accompanied by a more rapid spatial decay of projected TBDM\@. In contrast, the symmetric configuration maintains larger values at intermediate to long distances than in the mass asymmetric case. The power-law fits yield exponents of $\alpha = 1.42(2)$ for the symmetric system and $\alpha = 1.63(3)$ for the mass-asymmetric case, indicating faster decay in the latter over the fitting interval. These results show that mass asymmetry affects both the short-range e-h pairing amplitude and the long-distance behavior of e-h pair correlations. The projected TBDM obtained from VMC and DMC for the three representative regimes are in close agreement; see Appendix~\ref{Appendix G}. The energy scaling with system size from VMC and DMC for $N = 21$ to $N = 61$ fermions of each species is shown in Appendix~\ref{Appendix H}.  

\section{Conclusions}
In summary, we have employed diffusion quantum Monte Carlo to investigate the ground-state correlations of a mass-asymmetric 1D electron-hole biwire over a broad range of carrier densities and interwire separations. We identify three characteristic correlation regimes based on the combined behavior of PCFs, SSFs and the projected TBDM:\ a two-component plasma regime at high densities and relatively large interwire separations, an excitonic regime characterized by enhanced e-h correlations, and a low-density Wigner-like regime dominated by spatial ordering.

The excitonic regime is characterized by a short-range enhancement of the e-h PCF together with enhanced finite-size pair coherence signals in the projected TBDM\@. The power-law analysis characterizes the long-distance decay of the TBDM, consistent with the absence of true long-range condensate order in 1D\@.  Within the parameter range investigated strong excitonic correlations and finite-size long-distance pair coherence signals persist to high densities, reaching $r_{\rm s} = 0.3$ a.u.\ at $d = 0.01$ a.u. These results highlight the role of reduced dimensionality and interwire attraction in promoting e-h pairing.

We further find that mass-asymmetry affects the spatial character of excitonic correlations. For the representative case studied, increasing the hole-to-electron mass ratio from $\sigma = 1$ to $\sigma = 7$ enhances the short-range e-h pairing amplitude, while the larger fitted power-law exponent indicates faster decay of the projected TBDM over the fitting interval. Overall, our results provide DMC benchmarks for understanding the interplay of e-h pairing, algebraically decaying pair correlations, and spatial ordering in mass-asymmetric EHBWs.

\begin{acknowledgments}
G.S.\ acknowledges financial support from the Ministry of Education for this research. V.A.\ thanks the Science and Engineering Research Board, Anusandhan-National Research Foundation, Government of India, for funding under Core Research Grant No.\ CRG/2023/001573.
 We also thank the National Supercomputing Mission (NSM) for access to the PARAM Siddhi-AI supercomputing facilities at the National PARAM Supercomputing Facility (NPSF), C-DAC, Pune. The authors also acknowledge computational support from PARAM Shavak (`Gryphon') installed in V.A.'s laboratory at NIT Jalandhar developed by C-DAC\@.
\end{acknowledgments}

\section*{Data availability}
 The data that support the findings of this article are openly available in Ref.~\onlinecite{Shah2026Data}.

\appendix
\section{Finite-size errors}\label{Appendix A}

Finite-size effects were examined by performing simulations for an EHBW system with \(N=21\) particles of each species (42 particles total) and comparing selected observables with results obtained for a larger system size of \(N=61\) (122 particles total).  

\subsection{Projected Two-body density matrix}
\begin{figure}[tbp!]
	\centering
	
	\includegraphics[width=0.95\linewidth]{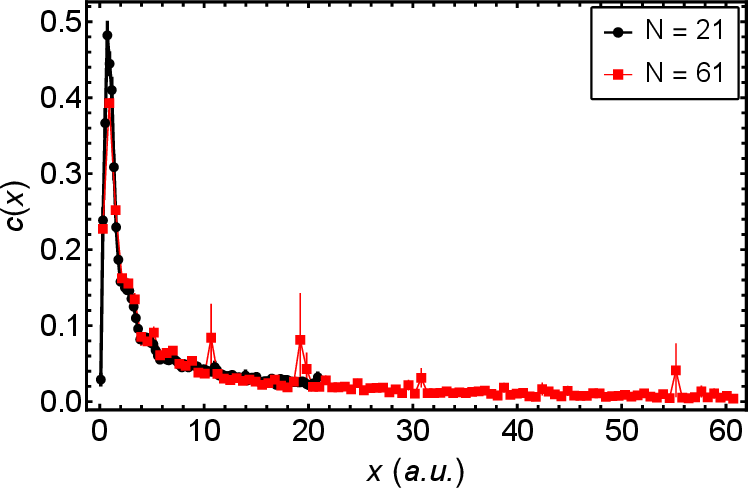}
	\caption{VMC projected TBDM $c(x)$ for an asymmetric EHBW system at $r_{\rm s} = 1.0$ a.u.\ and interwire separation $d = 0.07$ a.u. VMC results for $N=21$ are compared with $N=61$ particles of each species.}\label{fig:cf}
	
\end{figure}

Figure~\ref{fig:cf} shows the projected TBDM $c(x)$ for the EHBW system obtained from VMC simulations for $N = 21$ and $N = 61$ particles of each species. The increase in particle number extends the spatial range of the correlation function. Within the spatial region common to both system sizes, the overall structure of $c(x)$ remains quantitatively similar. The power-law fits yield exponents of $\alpha = 1.59(3)$ for $N = 21$ and $\alpha = 1.97(37)$ for $N = 61$. These results characterize the algebraic decay of the TBDM over the fitting interval for the two system sizes.

\subsection{Pair correlation function}

\begin{figure}[tbp!]
	\centering
		\includegraphics[width=0.95\linewidth]{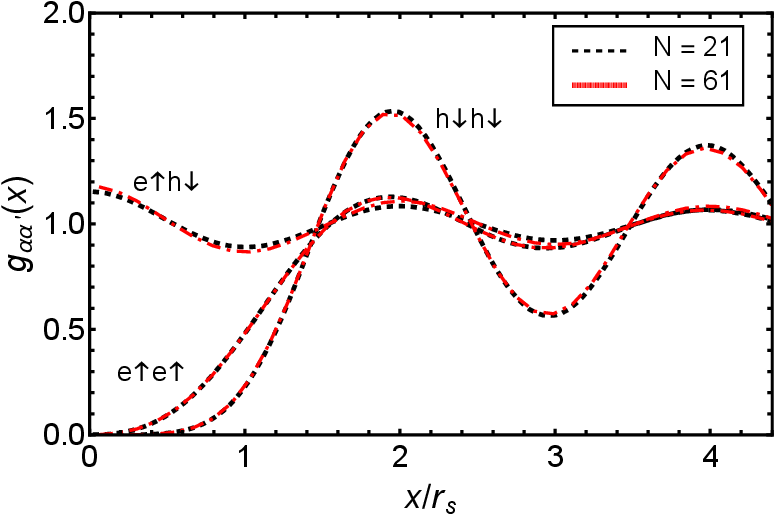}
		\caption{DMC PCFs for an asymmetric EHBW at $r_{\rm s} = 2.0$~a.u.\ and interwire separation $d = 2.0$~a.u. The system is in the two component fluid regime, and results are shown for system sizes $N = 21$ and $N = 61$.}\label{pcfp}
\end{figure}

\begin{figure}[tbp!]
		\centering
		\includegraphics[width=0.95\linewidth]{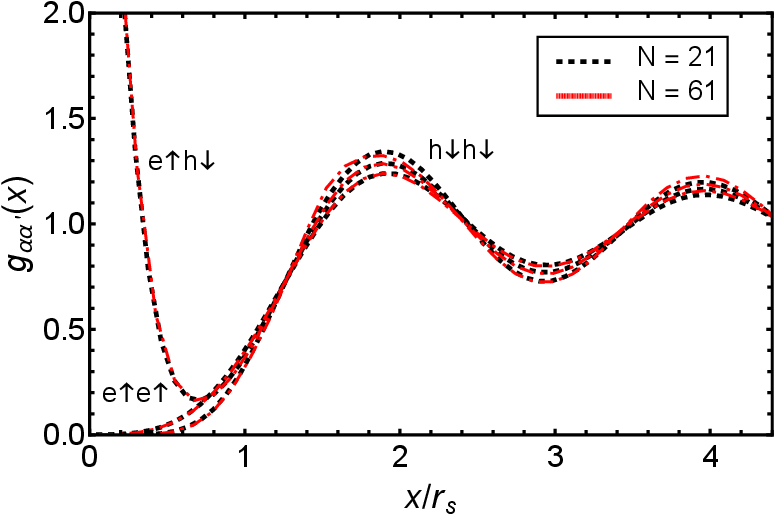}
		\caption{DMC PCFs for an asymmetric EHBW at $r_{\rm s} = 5.0$~a.u.\ and interwire separation $d = 1.0$~a.u. The system is in the excitonic regime, and results are shown for system sizes $N = 21$ and $N = 61$.}\label{pcfe}
\end{figure}

\begin{figure}[tbp!]
        \centering
	\includegraphics[width=0.95\linewidth]{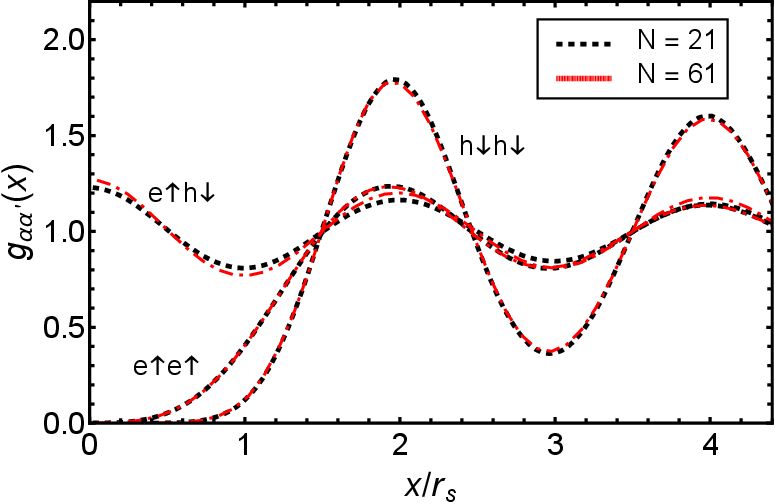}
	\caption{DMC PCFs for an asymmetric EHBW at $r_{\rm s} = 4.0$~a.u.\ and interwire separation $d = 5.0$~a.u. The system is in the Wigner ordering regime, and results are shown for system sizes $N = 21$ and $N = 61$.}\label{pcfw}
\end{figure}

Finite-size effects in the spin-resolved PCFs are summarized in Figs.~\ref{pcfp},~\ref{pcfe}, and~\ref{pcfw} for all the three regimes, showing excellent agreement between the two system sizes.
Any residual differences are well within the statistical uncertainties of the DMC data and display no systematic trend with \(N\). These observations indicate that finite-size effects are negligible in PCFs, and we therefore regard the \(N=21\) results as effectively converged with respect to system size.

\section{Long-distance power-law analysis} \label{Appendix B}

The resulting fit exponents for all three density regimes are
summarized in Table~\ref{powerlaw}.

\begin{table}[t]
	\caption{Long-distance power-law exponents $\alpha$ obtained
		from fits to the DMC projected two-body density matrix $c(x)$ for $N=21$
		particles of each species.}
	\label{powerlaw}
	\begin{ruledtabular}
		\begin{tabular}{ccc}
			$r_{\rm s}$ & $d$ & $\alpha$ \\
			\hline
			0.5 & 0.01 & 1.20(1) \\
			0.5 & 0.02 & 1.63(3) \\
			0.5 & 0.03 & 1.75(4) \\
			0.5 & 0.05 & 1.64(7) \\
			0.5 & 0.07 & 1.72(10) \\
			\hline
			1.0 & 0.05 & 1.47(1) \\
			1.0 & 0.07 & 1.59(2) \\
			1.0 & 0.09 & 1.48(2) \\
			1.0 & 0.10 & 1.45(3) \\
			1.0 & 0.30 & 1.76(10) \\
			1.0 & 0.50 & 1.80(18) \\
			\hline
			5.0 & 0.50 & 1.01(3) \\
			5.0 & 0.70 & 1.41(1) \\
			5.0 & 0.90 & 1.76(3) \\
			5.0 & 1.00 & 1.63(3) \\
			5.0 & 3.00 & 2.82(16) \\
			5.0 & 5.00 & 3.53(37) \\
		\end{tabular}
	\end{ruledtabular}
\end{table}

\section{Time-step errors}\label{Appendix C}

\begin{figure}[tbp!]
	\centering
	\includegraphics[width=0.95\linewidth]{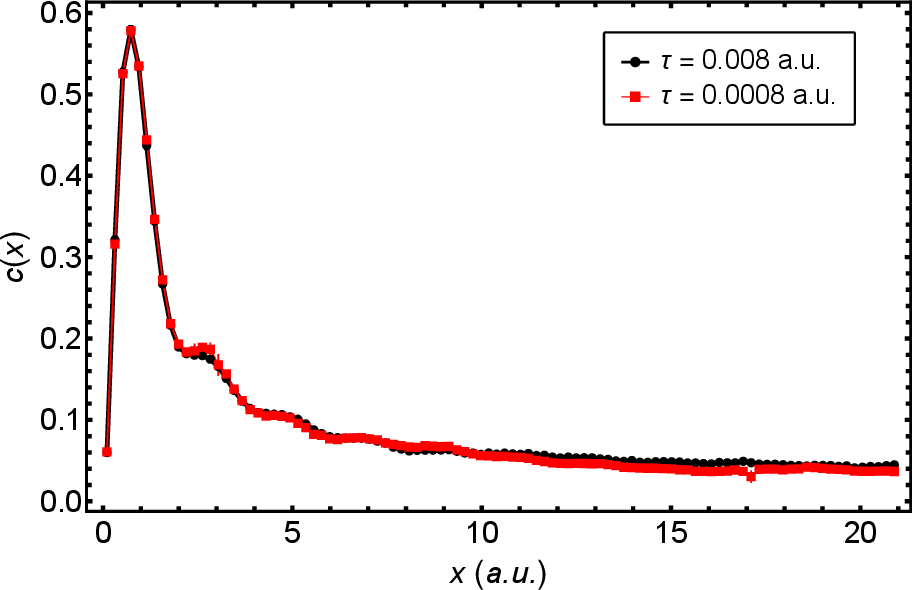}
	\caption{DMC projected TBDM $c(x)$ for an asymmetric EHBW system at $r_{\rm s} = 1.0$~a.u.\ and interwire separation $d = 0.05$~a.u. Results are shown for DMC time steps $\tau = 0.008$~a.u.\ and $\tau = 0.0008$~a.u.}\label{fig:cf_1}
\end{figure}

\begin{figure}[tbp!]
	\centering
	\includegraphics[width=0.95\linewidth]{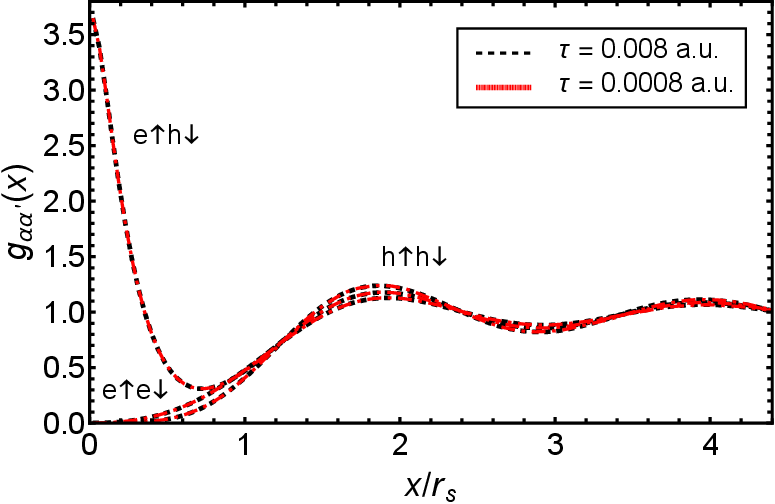}
	\caption{DMC PCFs for an asymmetric EHBW system at $r_{\rm s} = 1.0$~a.u.\ and interwire separation $d = 0.05$~a.u. Results are shown for DMC time steps $\tau = 0.008$~a.u.\ and $\tau = 0.0008$~a.u.}\label{fig:pcf_1}
\end{figure}

As discussed in the main text, an optimized time step 
was employed in the DMC calculations. This conservative choice was adopted to ensure that time-step bias is negligible across the entire parameter range studied. To explicitly verify this assumption, additional DMC simulations were performed at \(r_{\rm s} = 1\)~a.u.\ and \(d = 0.05\)~a.u.\ using a smaller time step of \(\tau = 0.0008\)~a.u., and the resulting expectation values were compared with those obtained at \(\tau = 0.008\)~a.u., as shown in Figs.~\ref{fig:cf_1} and~\ref{fig:pcf_1}. No statistically significant differences are observed between the two time steps. Since time-step errors are expected to be most severe in the high-density regime, we conclude that the results are well converged with respect to \(\tau\).

\section{Static structure factor}\label{Appendix D}

The spin-resolved SSF of a 1D EHBW is expressed as
\begin{align}
		S_{\alpha\alpha'}(k)
		=& \frac{N_\alpha}{N_{\rm e}+N_{\rm h}} \delta_{\alpha\alpha'} \nonumber \\ &+\frac{N_\alpha N_{\alpha'}}{(N_{\rm e}+N_{\rm h})L}
		\int_{-L/2}^{L/2} \bigl[g_{\alpha,\alpha'}(x)-1\bigr]
		e^{ik\cdot x}\ dx,
\end{align}
where $N_{\rm e}$ and $N_{\rm h}$ denote the numbers of electrons and holes, respectively (with $N_{\rm e}$ = $N_{\rm h}=N$ in the present case) and $g_{\alpha,\alpha'}(x)$ is the spin-resolved PCF\@.  

\begin{figure}[tbp!]
	\centering
		\includegraphics[width=0.95\linewidth,clip]{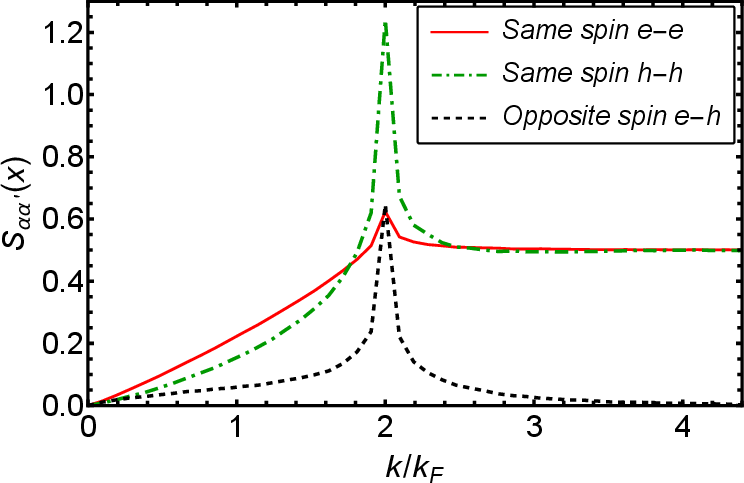}
		\caption{DMC estimate of the spin-resolved SSFs for an asymmetric EHBW in the two component fluid regime at $r_{\rm s} = 1.0$~a.u.\ and interwire separation $d = 0.5$~a.u.}\label{ssf_2c}
\end{figure}

\begin{figure}[tbp!]
		\centering
		\includegraphics[width=0.95\linewidth,clip]{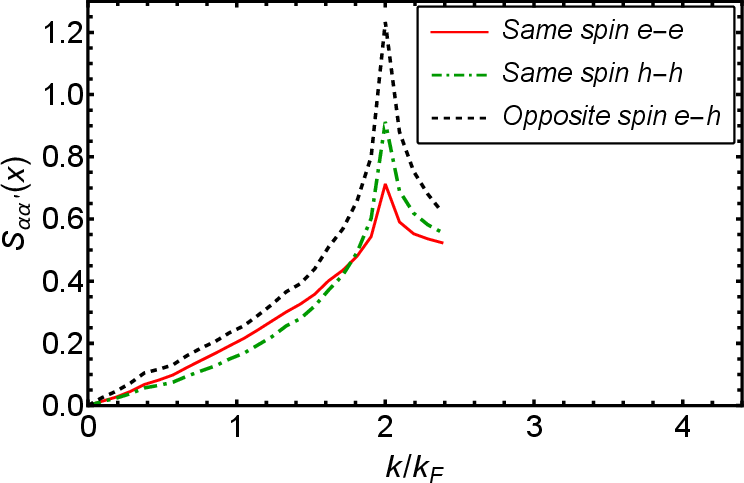}
		\caption{DMC estimate of the spin-resolved SSFs for an asymmetric EHBW in the excitonic regime at $r_{\rm s} = 0.5$~a.u.\ and interwire separation $d = 0.01$~a.u.}\label{ssf_ex} 
\end{figure}

\begin{figure}[tbp!]
  \centering
	\includegraphics[width=0.95\linewidth,clip]{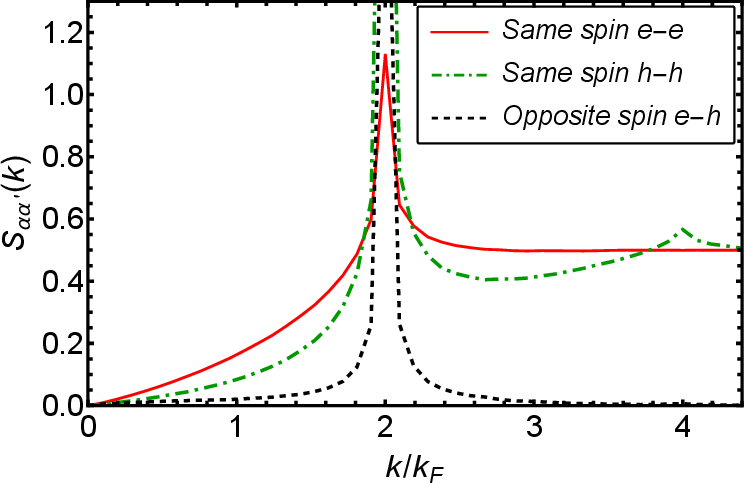}
	\caption{DMC estimate of the spin-resolved SSFs for an asymmetric EHBW in the Wigner-like regime at $r_{\rm s} = 5.0$~a.u.\ and interwire separation $d = 6.0$~a.u.}\label{ssf_wc}
\end{figure}

Figures~\ref{ssf_2c}--\ref{ssf_wc} show the DMC results for the spin-resolved SSF $S_{\alpha\alpha}(k)$ of the asymmetric EHBW system across three correlation regimes. The momentum dependence clearly illustrates the evolution from a 2C fluid to an excitonic regime and finally to a Wigner-correlated state.

In the fluid regime (Fig.~\ref{ssf_2c}), all components display the expected linear behavior in the long-wavelength limit. A cusp-like enhancement develops near $k \simeq 2k_{\rm F}$, indicating strengthened $2k_{\rm F}$ density correlations. The opposite-spin electron-hole (e-h) channel remains comparatively suppressed, demonstrating that while interwire Coulomb correlations are present, bound excitonic correlations are not dominant. The stronger $2k_{\rm F}$ feature in the hole-hole (h-h) component reflects the mass asymmetry and the enhanced correlations of the heavier species. In the excitonic regime (Fig.~\ref{ssf_ex}, obtained at $r_{\rm s} = 0.5$ a.u.\ and reduced interwire separation $d = 0.01$ a.u.), the opposite-spin e-h SSF becomes strongly enhanced around $k \approx 2k_{\rm F}$, exceeding the same-spin components. This pronounced peak signals dominant interwire pairing correlations and the formation of bound e-h pairs. The simultaneous enhancement of all channels near $2k_{\rm F}$, with the e-h component largest, indicates the buildup of interwire coherence and density-wave correlations driven by excitonic binding. In the Wigner-like regime (Fig.~\ref{ssf_wc}, at large $r_{\rm s} = 5.0$ a.u.), the $2k_{\rm F}$ feature sharpens dramatically and reaches its largest amplitude. The opposite-spin e-h channel exhibits a very narrow peak, while the same-spin (e-e/h-h) components show strong enhancement, with the h-h channel remaining slightly more pronounced due to the larger effective mass of the holes. A $4k_{\rm F}$ peak is also visible in the h-h SSF, providing an additional indication of spatial ordering in the heavier species. The redistribution of spectral weight toward $2k_{\rm F}$ together with the pronounced $4k_{\rm F}$ feature in the h-h SSF, is consistent with quasi-long-range crystalline order characteristic of strongly interacting 1D systems. In this regime, Coulomb repulsion and mass asymmetry dominate, driving the system toward a Wigner-like state with strong interwire density modulations.

\section{Momentum density}\label{Appendix E}
The momentum density (MD) for an EHBW system evaluated through the trial wave function $\psi$ is expressed as

\begin{equation}
	n_{\alpha}(k)=
	\frac{
		\int d\mathbf{R}\,|\Psi(\mathbf{R})|^2
		\sum_{i\in\alpha}
		\left[
		\int dx\,
		\frac{
			\Psi(x,x_2,\ldots,x_N)
		}{
			\Psi(x_{i,\alpha},x_2,\ldots,x_N)
		}
		\,e^{ik(x_{i,\alpha}-x)}
		\right]
	}
	{
		2 \pi N_{\alpha}
		\int d\mathbf{R}\,|\Psi(\mathbf{R})|^2
	},
\end{equation}
where $\alpha$ labels the particle species (electrons or holes), while the index $i \in \alpha$ runs over all particles of that species. The MD $n(k)$ encodes the occupancy of single-particle states in momentum space and is given by the Fourier transform of the one-particle density matrix. For a noninteracting Fermi gas,  $n(k)$ exhibits a discontinuity at the Fermi wave vector $k_{\rm F}$, while in the presence of interactions, the momentum distribution becomes continuous at $k_{\rm F}$ with a power-law singularity. In the present study, for an asymmetric EHBW system, the momentum distributions $n_\alpha(k)$ of electrons and holes differ, reflecting the underlying mass imbalance. We compute $n_\alpha(k)$ across a range of densities $r_{\rm s}$ and interwire separations $d$, encompassing all three characteristic regimes of the system.

\begin{figure}[tbp!]
	\centering
		\includegraphics[width=0.95\linewidth]{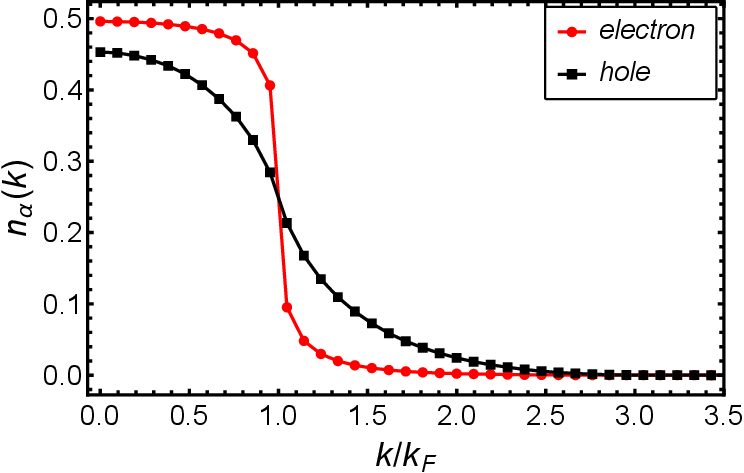}
		\caption{DMC MDs for an asymmetric EHBW system at $r_{\rm s} = 2.0$~a.u.\ and interwire separation $d = 2.0$~a.u. The system is in the two-component fluid regime.}\label{md_2c}
\end{figure}

\begin{figure}[tbp!]
		\centering
		\includegraphics[width=0.95\linewidth]{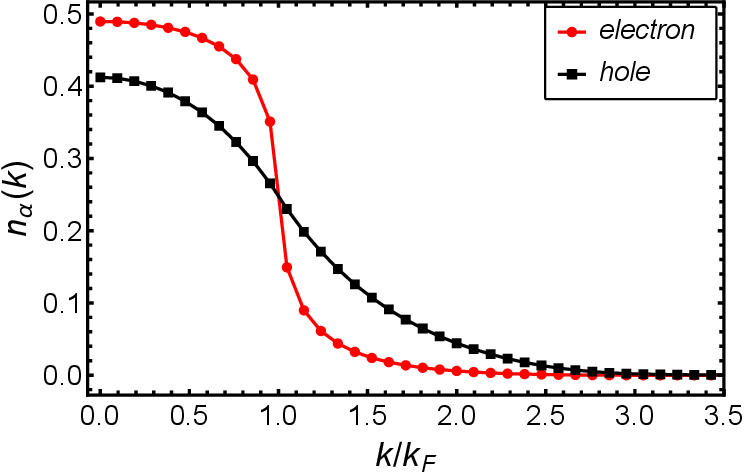}
		\caption{DMC MDs for an asymmetric EHBW system at $r_{\rm s} = 4.0$~a.u.\ and interwire separation $d = 5.0$~a.u. The system is in the Wigner-like regime.}\label{md_wc}
\end{figure}

\begin{figure}[tbp!]
	\centering
	\includegraphics[width=0.95\linewidth]{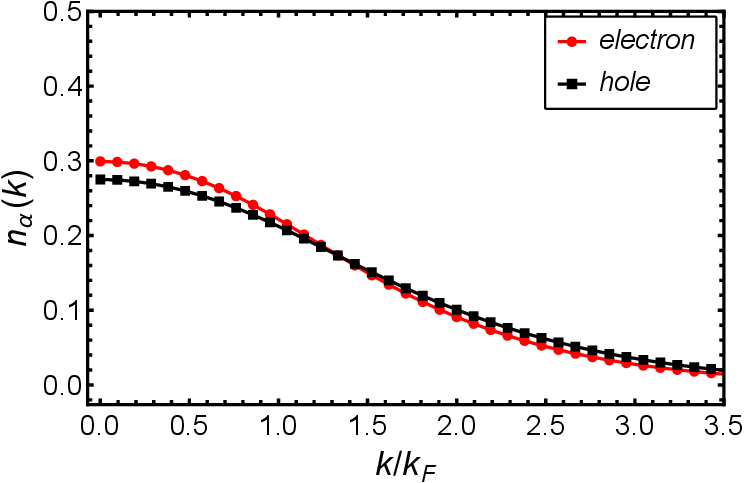}
	\caption{DMC MDs for an asymmetric EHBW system at $r_{\rm s} = 5.0$~a.u.\ and interwire separation $d = 1.0$~a.u. The system is in the excitonic regime.}
	\label{md_exciton}
\end{figure}

Figure~\ref{md_2c} shows the DMC-calculated $n(k)$ for a coupling parameter of $r_{\rm s} = 2.0$ a.u.\ and wire separation of $d = 2.0 $ a.u. The electron momentum distribution $n_{\rm e}$ exhibits a sharp drop at $k_{\rm F}$ whereas the hole momentum distribution $n_{\rm h}$ is broader around it because of its heavier mass. Nevertheless, both species retain recognizable fermionic momentum signatures, indicating that electrons and holes behave as two coupled but distinguishable fermionic components. At larger $r_{\rm s}$, the  interaction increases, while the increased interwire separation $d$ ($r_{\rm s} = 4.0$ a.u., $d = 5.0$ a.u.)\
weakens the electron-hole attraction. Under such conditions, the Coulomb interaction drives the system towards the Wigner-like state shown in Fig.~\ref{md_wc}. In this regime, the electrons and holes momentum distributions at the Fermi-wave vector
has different power-law features. Both $n_{e}(k)$ and $n_{h}(k)$ decrease smoothly over an extended range of momentum around $k \approx k_{\rm F}$. The hole $n(k)$ distribution remains broader compared to electrons, because of its larger mass, reflecting the strong spatial localization associated with Wigner-like ordering. When the interwire separation $d$ is reduced at a particular $r_{\rm s}$, the interwire interaction increases, leading to the formation of e-h bound pairs. In this regime ($r_{\rm s} = 5.0$ a.u., $d = 1.0$ a.u.), the fundamental degrees of freedom are no longer individual electrons and holes but correlated excitons as reflected in Fig.~\ref{md_exciton}. Both species $n(k)$
become nearly identical over the entire range, indicating strong pairing that locks the two species together. Furthermore, the distributions are smooth and show power law behavior. The close overlap of the $n(k)$ in both species provides compelling evidence for the formation of an excitonic regime.
\section{Two-particle density matrix in reciprocal space}\label{Appendix F}

The momentum space two-particle density matrix $\rho_{\alpha\alpha'}(k)$ provides a probe of pairing correlations and long-distance pair coherence in the asymmetric EHBW system.

\begin{figure}[tbp!]
	\centering
		\includegraphics[width=0.95\linewidth]{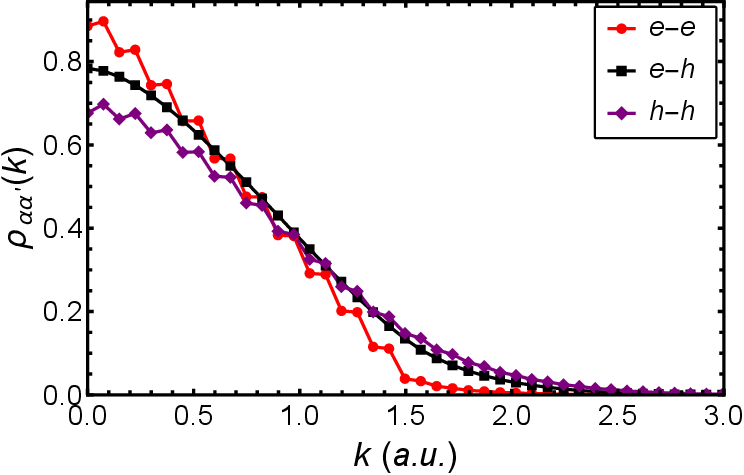}
		\caption{TBDM in momentum space for an asymmetric EHBW system at $r_{\rm s} = 2.0$~a.u.\ and interwire separation $d = 2.0$~a.u. The system is in the two-component fluid regime.}\label{tmd_2c}
\end{figure}

\begin{figure}[tbp!]
		\centering
		\includegraphics[width=0.95\linewidth]{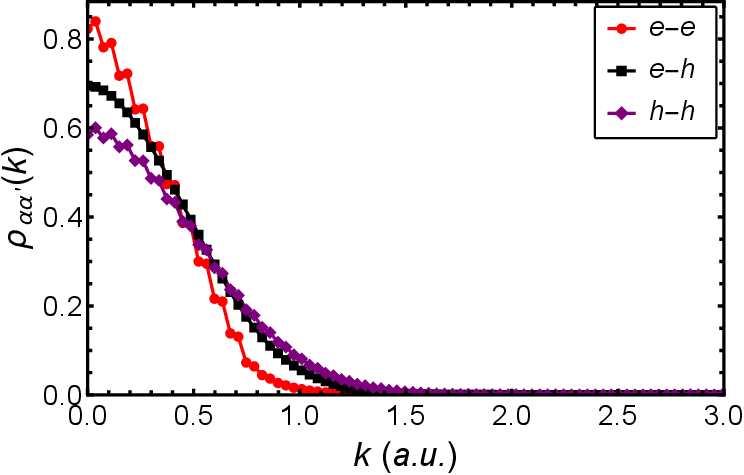}
		\caption{TBDM in momentum space for an asymmetric EHBW system at $r_{\rm s} = 4.0$~a.u.\ and interwire separation $d = 5.0$~a.u. The system is in the Wigner-like regime.}\label{tmd_wc}
\end{figure}
	
\begin{figure}[tbp!]
	\centering
	\includegraphics[width=0.95\linewidth]{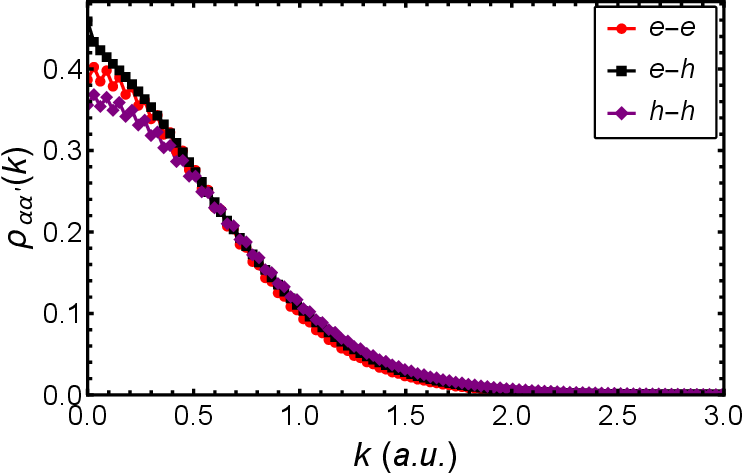}
	\caption{TBDM in momentum space for an asymmetric EHBW system at $r_{\rm s} = 5.0$~a.u.\ and interwire separation $d = 1.0$~a.u. The system is in the excitonic regime.}\label{tmd_exciton}
\end{figure}

For $r_{\rm s} = 2.0$ a.u.\ and separation $d = 2.0$ a.u.\ (Fig.~\ref{tmd_2c}), all spin-resolved components exhibit finite weight at small $k$, with gradual decay toward zero at larger momenta. The comparable magnitudes of the intra- and interspecies correlations indicate
that the system remains in a two-component correlated fluid regime. While electron-hole correlations are present, they do not dominate at the low 
$k$ response, signaling the absence of robust excitonic coherence. At density $r_{\rm s} = 4.0$ a.u.\ with increased $d = 5.0$ a.u.\ (Fig.~\ref{tmd_wc}), the overall magnitude of $\rho_{\alpha\alpha'}$(k) is reduced and the decay with momentum becomes sharper. The interwire correlations are suppressed due to enhanced spatial separation consistent with the Wigner-like regime, where carriers become spatially localized and interspecies coherence is energetically unfavorable. In contrast, at lower density and small separation ($r_{\rm s} = 5.0$ a.u.\ and $d = 1.0$ a.u., shown in Fig.~\ref{tmd_exciton}), the behavior changes. The electron-hole component is selectively enhanced at small $k$, exceeding the corresponding intrawire contributions. This redistribution of spectral weight toward the interspecies channel signals the formation of e-h bound pairs. The pronounced low-momentum amplitude reflects the buildup of long-wavelength 
coherence and is consistent with excitonic ordering.

\section{Comparison of VMC and DMC projected TBDM}\label{Appendix G}

\begin{figure}[tbp!]
	\centering
		\includegraphics[width=0.95\linewidth]{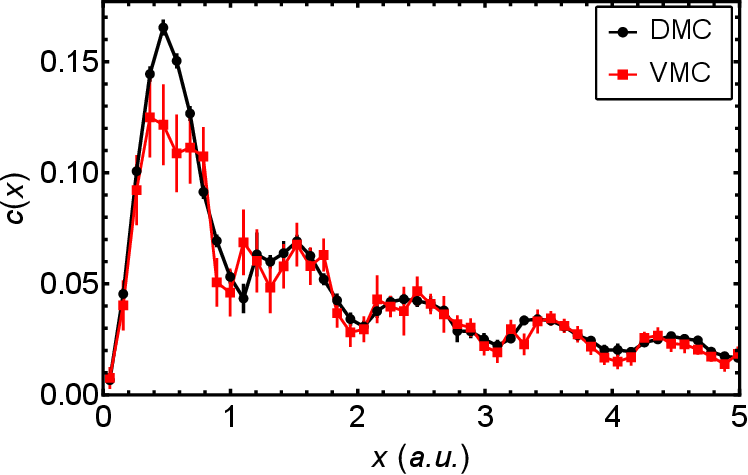}
		\caption{Comparison of VMC and DMC results for the projected TBDM $c(x)$ in the two-component fluid regime at $r_{\rm s} = 0.5$ a.u.\ and $d = 0.08$ a.u.}\label{vd_rs0.5}
\end{figure}

\begin{figure}[tbp!]
		\centering
		\includegraphics[width=0.95\linewidth]{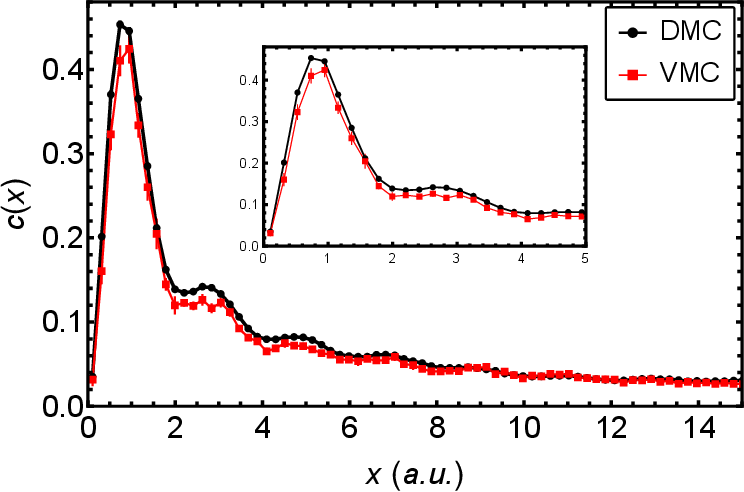}
		\caption{Comparison of VMC and DMC results for the projected TBDM $c(x)$ in the excitonic regime at $r_{\rm s} = 1.0$ a.u.\ and $d = 0.09$ a.u.}\label{vd_rs1} 
\end{figure}

\begin{figure}[tbp!]
	\centering
	\includegraphics[width=0.95\linewidth]{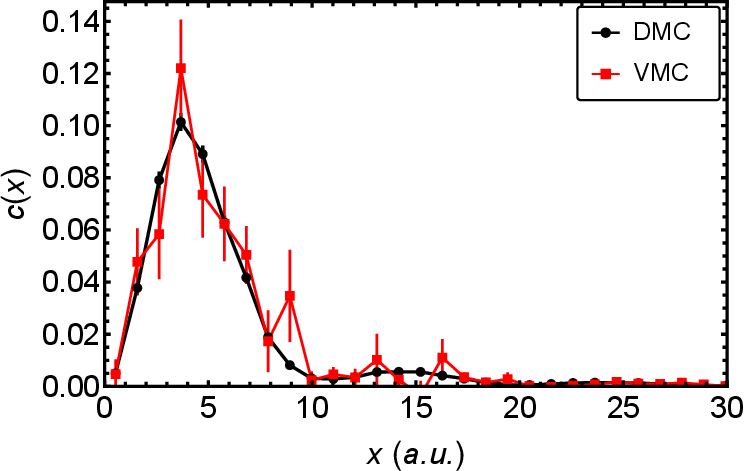}
	\caption{Comparison of VMC and DMC results for the projected TBDM $c(x)$ in the Wigner-like regime at $r_{\rm s} = 5.0$ a.u.\ and $d = 5.0$ a.u.}\label{vd_rs5}
\end{figure}

To assess the robustness of the quality of the trial wave function, we compare VMC and DMC results for the projected TBDM $c(x)$ across the distinct correlation regimes in this work.
In both VMC and DMC the error in the projected TBDM is first order in the error in the trial wave function, and there is no general way of combining the estimates to eliminate leading-order errors.
Expectation values are computed using the VMC method via importance-sampled Monte Carlo integration, where the sampling probability distribution is given by $|\Psi_{\rm T}|^2$, with $\Psi_{\rm T}$ denoting an appropriately chosen trial wave function. The trial wave function contains several variational parameters, which are independently optimized for each set of $d$ and $r_{\rm s}$ values. In the weak-coupling two-component fluid regime ($r_{\rm s} = 0.5$ a.u., $d = 0.08$ a.u.\ as in Fig.~\ref{vd_rs0.5}), both methods exhibit a pronounced peak followed by damped oscillations and a gradual decay. The DMC estimator tends to yield a slightly higher first maximum than VMC, while preserving the overall oscillatory structure. This indicates that the variational Ansatz captures the qualitative information, but slightly underestimates the strength of coherence. In the intermediate regime
($r_{\rm s} = 1.0$ a.u., $d = 0.09$ a.u.\ as in Fig.~\ref{vd_rs1}), the projected TBDM is amplified. The first peak is larger than in the fluid regime, and the slower decay of $c(x)$ indicates enhanced long-distance pair correlations. In contrast, in the strongly correlated Wigner-like regime ($r_{\rm s} = 5.0$ a.u., $d = 5.0$ a.u.\ as in Fig.~\ref{vd_rs5}), both methods show a localized peak at finite $x$ followed by a rapid decay of $c(x)$ to nearly zero. 
This indicates that once strong spatial localization and charge ordering dominate, the projected TBDM becomes intrinsically suppressed and largely insensitive to whether VMC or DMC sampling is used.
The absence of a sustained long-range tail reflects the destruction of extended excitonic coherence in this limit.

\section{VMC/DMC ground-state energy scaling with system size in an EHBW system}\label{Appendix H}

VMC and DMC energies in EHBW systems with $N=21$ and $N=61$ and various values of $r_{\rm s}$ and $d$ are compared in Table~\ref{table:E}. 

\begin{table}[t]
	\caption{VMC/DMC ground-state energy per particle $E(N)$ for $N=21$ and $61$ fermions of each species in an EHBW system. The regime labels denote two-component plasma (2CP), Wigner-like regime (WR), and excitonic regime (ER)\@. }
	\label{table:E}
	\centering
	\scriptsize
	\setlength{\tabcolsep}{3pt}
	\begin{ruledtabular}
		\begin{tabular}{@{}ccc cccc@{}}
			$r_{\rm s}$ & $d$ & Regime &
			\multicolumn{2}{c}{$E(21)$ (a.u.\ per particle)} &
			\multicolumn{2}{c}{$E(61)$ (a.u.\ per particle)} \\
			& & &
			VMC & DMC & VMC & DMC \\
			\hline
			2.0 & 2.0 & 2CP
			& $-0.266323(5)$ & $-0.266699(6)$
			& $-0.265844(6)$ & $-0.266585(4)$ \\
			
			4.0 & 5.0 & WR
			& $-0.238778(2)$ & $-0.238912(1)$
			& $-0.238624(2)$ & $-0.238876(1)$ \\
			
			5.0 & 1.0 & ER
			& $-0.323971(4)$ & $-0.324226(2)$
			& $-0.323773(3)$ & $-0.324216(1)$ \\
		\end{tabular}
	\end{ruledtabular}
\end{table}


\end{document}